\documentclass[fleqn]{llncs}

\usepackage[all]{xy}
\usepackage{color}
\usepackage{url}

\renewcommand{\tt}{\ttfamily}
\newcommand{\codefont}{\small\tt}
\newcommand{\code}[1]{\mbox{\codefont{#1}}}
\newcommand{\ccode}[1]{``\code{#1}''}
\newcommand{\us}{\raise-.8ex\hbox{-}}

\newcommand{\cpp}{C\raise 0.5ex\hbox{\tiny ++}}

\usepackage{listings}
\lstnewenvironment{curry}{\lstset{literate={->}{{$\rightarrow{}\!\!\!$}}3}}{}
\newcommand{\listline}{\vrule width0pt depth1.5ex}

\begin{document}
\pagestyle{plain} 
\sloppy

\title{ICurry}
\author{
S. Antoy \inst{1}
M. Hanus \inst{2}
A. Jost \inst{1}
S. Libby \inst{1}
}
\institute{
Computer Science Dept., Portland State University, Oregon, U.S.A.\\
\and
Institut f\"ur Informatik, Kiel University, D-24098 Kiel, Germany
}

\maketitle

\begin{abstract}
  FlatCurry is a well-established intermediate representation
  of Curry programs used in compilers that translate Curry code
  into Prolog and Haskell code.  Some FlatCurry constructs have
  no direct translation into imperative code. These constructs must be
  each handled differently when translating
  Curry code into C, \cpp{} and Python code.
  We introduce a new representation of Curry programs, called ICurry,
  and derive a translation from all FlatCurry constructs into ICurry.
  We present the syntax of ICurry and the translation from FlatCurry to ICurry.
  We present a model of functional logic computations as graph rewriting,
  show how this model can be implemented in a low-level imperative language,
  and describe the translation from ICurry to this model.
\end{abstract}


\section{Introduction}
\label{Introduction}

Functional logic languages \cite{AntoyHanus10CACM}
provide fast software prototyping and development,
simple elegant solutions to otherwise complicated problems,
a tight integration between specifications and code
\cite{AntoyHanus12PADL},
and an ease of provability \cite{AntoyHanusLibby17EPTCS,Hanus18PPDP}
unmatched by other programming paradigms.
Not surprisingly, these advantages place heavy demands on their
implementation.  Theoretical results must be proven and efficient models of execution must be developed.
For these reasons,
the efficient implementation of functional logic languages
is an active area of research with contributions from many sources.
This paper is one such contribution.

Compilers of high-level languages transform a \emph{source} program
into a \emph{target} program which is in a lower-level language.
This transformation maps constructs available
in the source program language into simpler, more primitive, constructs
available in the target program language.
For example, pattern matching can be translated into a sequence
of \emph{switch} and assignment statements available in C, Rust, or Java.
We use this idea to map Curry into a C-like language.
Our target language is not standard C, but a more abstract language that
we call \emph{ICurry}.  The ``I'' in ICurry stands for ``imperative'',
since a design goal of the language is to be easily mappable into an
imperative language.

There are advantages in choosing ICurry over C.
ICurry is simpler than C.  It has no arrays, \emph{typedef}
declarations, types, explicit pointers, or dereferencing operations.
ICurry is more abstract than concrete low level languages.
Because of its simplicity and abstraction,
it has been mapped with a modest effort to C, \cpp{}, and Python.

Section~\ref{Curry} is a brief overview of Curry, with focus on the
features relevant to ICurry or to the examples.
Section~\ref{The Model} discusses an operational model of execution for
functional logic computations.  This model can be implemented
relatively easily in ICurry or in common imperative languages.
Section~\ref{FlatCurry} presents \emph{FlatCurry}, a format
of Curry programs similar to ICurry.  FlatCurry has been used
in the translation of Curry into other, non-imperative, languages, but it is not
suitable for the translation of Curry into an imperative language.
Section~\ref{ICurry} defines ICurry and
discusses its use and generation.
A detailed algorithm for obtaining
ICurry from FlatCurry is presented in Appendix A.
Section~\ref{Related Work and Conclusion} addresses related work
and offers our conclusion.

\section{Curry}
\label{Curry}

Curry is a declarative language
that joins the most appealing features of functional and
logic programming.
A \emph{Curry program} declares \emph{data types},
which describe how information is structured,
and defines \emph{functions} or \emph{operations},
which describe how information is manipulated.
For example:
\begin{curry}
  data List a = Nil | Cons a (List a)
\end{curry}
declares a polymorphic type \code{List} in which \code{a}
is a type parameter standing for the type of the list elements.
The symbols \code{Nil} and \code{Cons}
are the \emph{constructors} of \code{List}.
The values of a list are either \code{Nil}, the empty list,
or \code{Cons$\;e\;l$}, a pair in which $e$
is an element and $l$ is a list.

Since lists are ubiquitous, a special notation eases writing and
understanding them.  We use \code{[\,]} to denote the empty list
and $(e\,\code{:}\,l)$ to denote the pair.
A finite list is written \code{[$e_1$,$\ldots$,$e_n$]},
where $e_i$ is a list element.  For example,
\code{[1,2,3]} $=$ \code{1:2:3:[\,]}.

\emph{Functions} are defined by rewrite rules of the form:
\begin{equation}
  \label{rewrite-rule}
  \begin{array}{@{}l l @{}}
    f\; \bar p & \mid c_1 = e_1 \\
               & \cdots \\
               & \mid c_n = e_n 
  \end{array}
\end{equation}
where $f$ is a function symbol, $\bar p$ stands for a sequence
of zero or more expressions made up only of constructor symbols and variables,
``$|\; c_i$'' is a condition, and $e_i$ is an expression.
Conditions in rules are optional.
The expressions in $\bar p$ are called \emph{patterns}.
For example, consider:
\begin{equation}
  \label{length-rules}
\vbox { \hbox {\hspace*{-2em}
\begin{curry}
  abs x | x <  0 = -x
        | x >= 0 =  x  $\listline$
  length []     = 0
  length (_:xs) = 1 + length xs
\end{curry}
}}
\end{equation}
where
\code{abs} computes the absolute value of its argument
and shows some conditions, and
\code{length} computes the length of its argument
and shows some patterns.

In contrast to most other languages, the textual order of the
rewrite rules in a program is irrelevant---all the rules
that can be applied to an expression are applied.
An emblematic example is a function, called \emph{choice},
and denoted by the infix operator \ccode{?}, which chooses
between two \emph{alternatives}:
\begin{curry}
  x ? y = x
  x ? y = y
\end{curry}
Therefore, \code{0\,?\,1} is an expression that produces \code{0}
and \code{1} non-deterministically.
In Curry, there are many other useful syntactic and semantic features,
for example, rewrite rules can have nested scopes with local definitions.
We omit their description here, since they are largely irrelevant
to our discussion, with the exception of \emph{let blocks} and
\emph{free variables}.

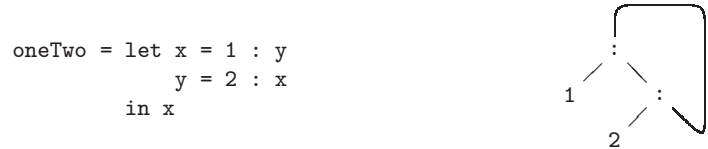
\begin{figure}[ht]
  \vspace*{-.25in}
  \hspace*{0.5in}
\begin{minipage}{2in}
  \tt
\begin{tabbing}
  oneTwo = \= let \= x = 1 : y \\
  \> \> y = 2 : x \\
  \> in x
\end{tabbing}
\end{minipage}
  \hspace*{1in}
\begin{minipage}{2in}
\[
\begin{xy}
0;<0.6mm,0mm>:  
(10,0)*+{\mbox{\tt 2}}="2";
(0,10)*+{\mbox{\tt 1}}="1";
(20,10)*+{\mbox {\tt :}}="lc";
(10,20)*+{\mbox{\tt :}}="uc";
{\ar@{-} "uc";"lc"};
{\ar@{-} "uc";"1"};
{\ar@{-} "lc";"2"};
**\crv{(20,10)&(25,5)&(30,0)&(30,5)&(30,25)&(30,30)&(25,30)&(15,30)&(10,30)&(10,25)&(10,20)}
\end{xy}
\]
\end{minipage}
\caption{\label{let-example-graph}
  Example of a let block with mutually recursive variables
  and the graph it defines.
}
\end{figure}

Let blocks support the definition of circular expressions
which allows the construction of cyclic graphs.
Fig.~\ref{let-example-graph} shows an example of a let block
and the corresponding graph.
Expression \code{oneTwo} evaluates to the
infinite list \code{1:2:1:2:$\ldots$}
\newcommand{\xcirc}{}

Free variables abstract unknown information and are
``computationally inert'' until the information
they stand for is required during a computation.
When this happens, plausible values for a variable are
non-deterministically produced by narrowing
\cite{AntoyEchahedHanus00JACM,Reddy85}.
Free variables might occur in initial expressions, conditions,
and the right-hand side of rules, and need to be declared by the
keyword \code{free}, unless they are anonymous (denoted by \ccode{\us{}}).
For instance, the following program
defines list concatenation
which is exploited to define an operation that returns
some element of a list having at least two occurrences:
\begin{curry}
(++) :: [a] -> [a] -> [a]
[]     ++ ys = ys
(x:xs) ++ ys = x : (xs ++ ys)$\listline$
someDup :: [a] -> a
someDup xs | xs == _$\;$++$\;$[x]$\;$++$\;$_$\;$++$\;$[x]$\;$++$\;$_
           = x    where$\;$x$\;$free
\end{curry}

\section{The Execution Model}
\label{The Model}

A \emph{program} is a graph rewriting system
\cite{EchahedJanodet97GraphRewriting,Plump99Handbook}
over a \emph{signature}, partitioned into \emph{constructor} and \emph{operation} symbols.
We briefly and informally review the underlying theory.
A \emph{graph} is a set of \emph{nodes}, where a node is an object
with some attributes, and an identity by virtue of being an element in a set.
Key attributes of a node are a \emph{label} and a sequence of \emph{successors},
A label is either a symbol of the signature or a variable.
A successor is another node, and the sequence of successors may be empty.
Exactly one node of a graph is designated as the graph's \emph{root}.
Each node of a graph corresponds to an expression in the Curry program.

A \emph{graph rewriting system} is a set of rewrite rules
following the \emph{constructor discipline} \cite{Odonnell-85}.
A \emph{rule} is a pair of graphs, $l \to r$,
called the left- and right-hand sides respectively.
Rules are unconditional without loss of generality \cite{Antoy01PPDP}.
A \emph{rewrite step} of a graph $e$ 
first identifies both a subgraph $t$ of $e$, and
a rule $l \to r$ in which
$t$ is an instance of $l$, then replaces $t$ with
the corresponding instance of $r$.
The identification of the subgraph $t$ and the rule $l \to r$
is accomplished by a \emph{strategy} \cite{Antoy05JSC}.
For example, given the rules (\ref{length-rules}),
a step of \code{length [3,4]} produces \code{1+length[4]}
where the subgraph reduced in the step is the whole graph,
and the rule applied in the step is the second one.

\begin{figure}[t]
\vspace*{-3em}
\hspace*{1.0in}
\xymatrix@1@C=5pt@R=6pt@M=3pt{
  & \mbox{\tt +} \ar@{-}[ddl] \ar@{-}[ddr] \\ \\
  \mathtt{coin} & & \mathtt{coin} \\
  & \makebox[0pt]{\tt coin+coin}
}
\hspace{6em}
\xymatrix@1@C=5pt@R=6pt@M=3pt{
  & \mbox{\tt +} \ar@{-}@/_2ex/[dd] \ar@{-}@/^2ex/[dd] \\ \\
  & \mathtt{coin} \\
  & \makebox[0pt]{\tt x+x where x=coin}
}
\caption{\label{sharing-pict}
  Graphical and textual representation of expressions.
  In Curry, all the occurrences of the same variable are shared.
  Hence, the two occurrences of \code{x} stand for the same node.
  The expression \code{coin} is conventionally an integer constant with two values,
  \code{0} and \code{1}, non-deterministically chosen.
  The sets of values produced by the two expressions differ.
  }
\end{figure}
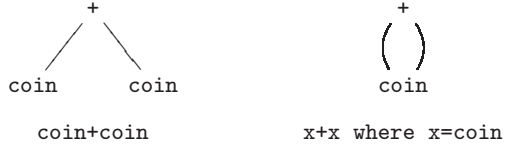

A \emph{computation} of an expression $e$
is a sequence of rewrite steps starting with $e$,
$e = e_0 \to e_1 \to \ldots$~
Expression $e$ is referred to as \emph{top-level}, and
each $e_i$ as a \emph{state} of the computation of $e$.
A \emph{value} of a computation is a state in which every node is labeled by
a constructor symbol.
Such expression is also called a \emph{constructor normal form}.
Not every computation has values.

We have modeled a functional logic program as a graph rewriting system
\cite{EchahedJanodet97GraphRewriting,Plump99Handbook}.
Functional logic computations are executed in this model
by rewriting which consists of two relatively simple operations:
the construction of graphs and the replacement of subgraphs
with other graphs.
The most challenging part is selecting the subgraph
to be replaced in a way that does not consume computational resources
unnecessarily.  This is a well-understood problem \cite{Antoy05JSC}
which is largely separated from the model.

In an implementation of the model,
the expressions are objects of a computation and are
represented by dynamically linked structures.
These structures are similar to those used for
computing with lists and trees.  The nodes of such a structure are in
a bijection with the nodes of the graph they represent.  Unless a
distinction is relevant, we do not distinguish between a graph and
its representation.

The occurrence of a symbol, or variable, in the textual representation
of an expression stands for the node labeled by the occurrence.
Distinct occurrences may stand for the same node,
in which case we say that the occurrences are \emph{shared}.
The textual representation accommodates this distinction,
therefore it is a convenient, linear notation for a graph.
Fig.~\ref{sharing-pict} shows two graphs and their corresponding
textual expressions.

\section{FlatCurry}
\label{FlatCurry}

\emph{FlatCurry} \cite{flatcurry} is an intermediate language used
in a variety of applications.
These applications include implementing Curry by compiling into other languages,
like Prolog \cite{Hanus18PAKCS} or Haskell \cite{BrasselHanusPeemoellerReck11}.
FlatCurry is also the basis for specifying the operational semantics
of Curry programs \cite{AlbertHanusHuchOliverVidal05},
building generic analysis tools \cite{HanusSkrlac14},
or verifying properties of Curry programs \cite{Hanus17LOPSTR,Hanus18PPDP}.
The FlatCurry format of a Curry program removes some syntactic constructs,
such as nested scopes and infix notation, 
that make source programs more human readable.
This removal still preserves the program's meaning.
We ignore some elements of FlatCurry,
such as imported modules or exported symbols, which are not
directly related to the execution model presented in
Section \ref{The Model}.
Instead, we focus on the declaration of data constructors, the
definition of functions, and the construction of expressions.
These are the elements that play a central role in our execution model.

FlatCurry is a machine representation of Curry programs.
As such, it is not intended to be read by human.
For example, each variable is identified by an integer,
there is only prefix function application, and
pattern matching is broken down into a cascade of case distinctions.
In the examples that follow,
we present a sugared version of FlatCurry in which
variables have symbolic names, typically the same as in Curry;
the application of familiar infix operators is infix;
and indentation, rather than parentheses and commas,
show structure and grouping.
The intent is to make the examples easier to read
without altering the essence of FlatCurry.

In FlatCurry, 
data constructors are introduced by a type declaration.
A type $t$ has attributes such as a name and a visibility,
and chief among these attributes is a set of constructors $c_1, c_2, \ldots c_n$.
Each constructor $c_i$ has similar attributes, along with an arity and type
of each argument, which are not explicitly used in our discussion.
The same information is available for operation symbols,
but operations contain their code.

\begin{figure}[t]
\[
\begin{array}{lcl@{~~~~}l}
D & ::= & f(x_1,\ldots,x_n) = e  & \mbox{(function definition)} \\
e & ::= & x & \mbox{(variable) } \\
  & | & c(e_1,\ldots,e_n) & \mbox{(constructor call) } \\
  & | & f(e_1,\ldots,e_n)  & \mbox{(function call) } \\
  & | & \mathit{case}~e~\mathit{of}~\{p_1\to e_1; \ldots; p_n \to e_n\}
                         & \mbox{(case expression) } \\
  & | & e_1~\mathit{or}~e_2 & \mbox{(disjunction) } \\
  & | & \mathit{let}~\{x_1 = e_1;\ldots;x_n = e_n\} ~\mathit{in}~ e
       & \mbox{(let binding) } \\
  & | & \mathit{let}~x_1,\ldots,x_n ~\mathit{free~ in}~ e
       & \mbox{(free variables) } \\
p & ::= & c(x_1,\ldots,x_n)     & \mbox{(pattern)} 
\end{array}
\]
\caption{Abstract syntax of function definitions in FlatCurry}
\label{fig:flatcurry}
\end{figure}

The abstract syntax of FlatCurry operations is summarized
in Fig.~\ref{fig:flatcurry}.\footnote{In contrast to some other
presentations of FlatCurry (e.g., \cite{AlbertHanusHuchOliverVidal05,Hanus13}),
we omit the difference between rigid and flexible case expressions.}
Each operation is defined by a single rule with a linear left-hand side,
i.e., the argument variables $x_1,\ldots,x_n$ are pairwise different.
The right-hand side of the definition consists of (1) variables
introduced by the left-hand side or by a \emph{let block}
or by a $\mathit{case}$ pattern,
(2) constructor or function calls,
(3) $\mathit{case}$ expressions, (4) disjunctions,
(5) $\mathit{let}$ bindings, or (6) introduction of free variables.
The patterns $p_i$ in a $\mathit{case}$ expression
must be pairwise different constructors applied to variables.
Therefore, deep patterns in source programs are
represented by nested $\mathit{case}$ expressions.

Case expressions closely resemble definitional trees.
We recall that a \emph{definitional tree} of some operation $f$,
of arity $n$, is a hierarchical structure of expressions of the form
$f\;p_1\;\ldots\;p_n$, where $p_i$ is a pattern.
Since $f$ is constant and provides no information,
except to ease readability, we also call these expressions patterns.
The pattern at the root of the tree is $f\;x_1\;\ldots\;x_n$,
where the $x_i$'s are distinct variables.
The patterns at the leaves
are the left-hand sides of the rules of $f$,
except from the names of the variables.
For ease of understanding, 
in pictorial representations of definitional trees
we add the right-hand side of the rules too.
If $f\;p_1\;\ldots\;p_n$ is a branch node, $\beta$, of the tree,
a variable $x$ in some $p_j$ is singled out.
We call the variable $x$ \emph{inductive}.
The pattern in a child of $\beta$ is $f\;p_1\;\ldots\;q_j\;\ldots\;p_n$
where $q_j$ is obtained from $p_j$ by replacing $x$ with
$c\;y_1\;\ldots\; y_k$, where $c$ is a constructor of the type of $x$
and each $y_i$ is a fresh variable.
For example, consider the usual operation \code{zip} for
zipping two lists:
\begin{equation}
  \label{zip-rules}
\vbox { \hbox {\hspace*{-2em}
\begin{curry}
  zip []      y       = []
  zip (x1:x2) []      = []
  zip (x1:x2) (y1:y2) = (x1,y1) : zip x2 y2
\end{curry}
}}
\end{equation}
The corresponding definitional tree is shown below
where the inductive variable is boxed.
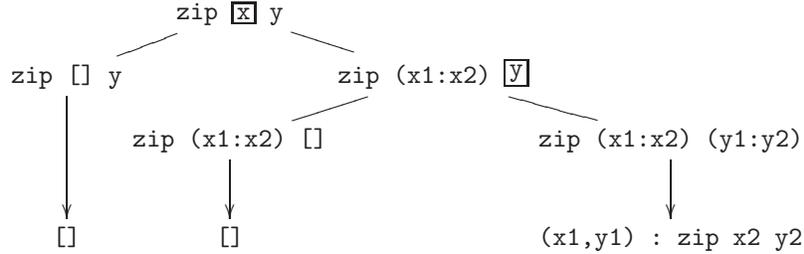
\newcommand{\boxit}[1]{\raisebox{-1.5pt}{\vbox{\hrule\hbox{\vrule\kern1.5pt
  \vbox{\kern1.5pt\hbox{#1}\kern1.5pt}\kern1.5pt\vrule}\hrule}}}
\begin{displaymath}
  \label{def-tree-example}
  \kern-1em
\xymatrix@1@C=-2pt@R=8pt@M=3pt{
  & \mbox{\tt zip \boxit{x} y}  \ar@{-}[dl] \ar@{-}[dr] \\
  \mbox{\tt zip [] y} \ar[ddd] 
  & & \mbox{\tt zip (x1:x2) \boxit{y}}  \ar@{-}[dl]\ar@{-}[dr] \\
  & \mbox{\tt zip (x1:x2) []} \ar[dd]  & & \mbox{\tt zip (x1:x2) (y1:y2)} \ar[dd] \\
  \\
  \mbox{\tt []} & \mbox{\tt []}  & & \mbox{\tt (x1,y1) : zip x2 y2}
}
\end{displaymath}
\noindent
The FlatCurry code of the rules of operation \code{zip},
closely corresponds to the code below.
This would be harder for the programmer to write than (\ref{zip-rules})
and less readable, but is semantically equivalent.
Every program can be transformed into an equivalent program in which
every operation has a definitional tree \cite{Antoy01PPDP}
which can be obtained from the operation's rules
with a relatively simple algorithm \cite{Antoy05JSC}.
\begin{curry}
  zip x y = case x of
              []      -> []
              (x1:x2) -> case y of
                            []      -> []
                            (y1:y2) -> (x1,y1) : zip x2 y2 
\end{curry}
Expressions are the final relevant element of FlatCurry.
As the code of \code{zip} shows, an expression can be a literal,
like \code{[]}; an application of constructors and operations
to expressions possibly containing variables, like
\code{(x1,y1)\,:\,zip x2 y2}; or a case expression,
like \code{case y of $\ldots$}~
FlatCurry also has \emph{let blocks}
to support the construction of cyclic graphs,
as shown in Fig.~\ref{let-example-graph}.

FlatCurry programs cannot be directly mapped to code in
a C-like target language.  There are two problems:
case expressions as arguments of a symbol application,
and let blocks with shared or mutually recursive variables.
A contrived example of the first is:
\begin{curry}
  3 + case x of True -> 1; False -> 2
\end{curry}
An example of the second is shown in Fig.~\ref{let-example-graph}.
ICurry proposes a solution to these problems
in a language-independent form which is suitable for the imperative paradigm.

\section{ICurry}
\label{ICurry}

In this section we define ICurry, 
discuss how to map it to imperative code that implements
our earlier model of computation,
and show how to obtain it from FlatCurry.

\subsection{ICurry Definition}

\emph{ICurry} is a format of Curry programs similar in intent to FlatCurry.
The purpose of both is to represent a Curry program into a format
with a small number of simple constructs.
Properties and manipulations of programs can be more easily investigated
and executed in these formats.
ICurry is specifically intended for compilation into a low-level language.
Each ICurry construct can be translated into 
a similar construct of languages such as C, Java or Python.
This should become apparent once we describe the constructs.

ICurry's data consists of nested applications of symbols represented as graphs.
ICurry's key constructs provide
the declaration or definition of symbols and
variables, construction of graph nodes, assignment,
and conditional executions of these constructs.
Rewriting steps are implemented in two phases,
once the redex and rule are determined. First, the replacement of the redex
is constructed. This is defined by the right-hand side of the rule.
Then, the successors pointing to the root of the redex
are redirected \cite[Def. 8]{EchahedJanodet97GraphRewriting},
through assignments, to point to the root
of the replacement.

The declaration of data constructors in ICurry is identical to that
in FlatCurry as described earlier.
However, the constructors of a type are in an arbitrary, but fixed, order.
Therefore, we can talk of the first, second, etc., constructor of a type.
This index is an attribute of constructor symbols which we call the \emph{tag}.
The tag is used to provide efficient pattern matching
and to simplify the execution of some strategy.
We will return to this topic in Sect.~\ref{ICurry Use}.

\begin{figure}[t]
\[
\begin{array}{lcl@{~~~~}l}
D & ::= & f = blck  & \mbox{(function definition)} \\
blck & ::= & decl_1 \ldots decl_k ~ asgn_1 \ldots asgn_n ~ stm & \mbox{(block) } \\
decl & ::= & \mathit{declare}~x & \mbox{(local variable declaration) } \\
     &  |  & \mathit{free}~x & \mbox{(free variable declaration) } \\
asgn & ::= & v = exp & \mbox{(variable assignment) } \\
stm  & ::= & \mathit{return}~ exp & \mbox{(return statement) } \\
     &  |  & \mathit{exempt} & \mbox{(failure statement) } \\
     & | & \mathit{case}~x~\mathit{of}~\{c_1\to blck_1; \ldots; c_n \to blck_n\}
                         & \mbox{(case statement) } \\
exp  & ::= & v & \mbox{(variable) } \\
     &  |  & \mathit{NODE}(l,exp_1,\ldots,exp_n) & \mbox{(node construction) } \\
      & | & exp_1~\mathit{or}~exp_2 & \mbox{(disjunction) } \\
v & ::= & x    & \mbox{(local variable)} \\
   & |  & v[i] & \mbox{(node access) } \\
   & |  & \mathit{ROOT} & \mbox{(root of function call)} \\
l & ::= & c    & \mbox{(constructor symbol)} \\
  &  |  & f    & \mbox{(function symbol)} \\
  &  |  & LABEL(v)  & \mbox{(node label symbol)} \\
\end{array}
\]
\caption{Abstract syntax of function definitions in ICurry}
\label{fig:icurry}
\end{figure}

The abstract syntax of operations in ICurry is summarized
in Fig.~\ref{fig:icurry}.
In FlatCurry, the body of a function is an expression.
In ICurry, it is a block of statements, the last of which returns an expression.
We describe expressions first.

Expressions are nested symbol applications represented as graphs.
Therefore, 
an expression is either a \emph{variable}, which refers to a node in a graph,
or a \emph{symbol application}.
ICurry makes an application explicit with a \emph{directive},
\code{NODE}, that constructs a graph node
from its label (1st argument) and its successors (remaining
arguments), and returns a reference to the node.
Accordingly, there are directives to access node components:
assuming $x$ is a variable referring to a node,
$\code{LABEL}(x)$ retrieves the label, and
$x[k]$ retrieves the $k$-th successor of the node.
A variable is either \emph{free} or \emph{local}.
The ICurry format distinguishes between constructor and
function application, and between full and partial application.
We do not discuss these details in this paper.
It is expected that by providing this additional information,
processors will be able to generate low-level code more easily,
and the generated code should be easier to optimize.

In ICurry, there are only a handful of statement kinds:
\emph{declaration} of a variable, and \emph{assignment} to a variable,
\emph{return}, and \emph{case} expressions.
Following FlatCurry, variables are represented by integers.
A declaration introduces a variable.
An assignment to a variable is a reference to a graph node.
Successors of a node referenced by $x$ are accessed through
the $x[\ldots]$ construct.
Arguments passed to functions are accessed through local variables.
The return statement is intended to return an expression,
the result of a function call.
Case expressions in ICurry are structurally similar to those
in FlatCurry, but with two differences for algebraically defined types,
which have a finite number of data constructors.
First, the branches of a case expression are in tag order.
We will justify this decision in Sect.~\ref{ICurry Use}.
Second, the set of branches of a case expression is complete,
i.e., there is a branch for each constructor of the type.

ICurry code begins with a declaration, and possible assignment, of some variables.
It is then followed by either a case statement, or a return statement.
Each branch of the case expression may declare and assign variables,
and may lead to either another case statement, or a return statement.

Below, we present two examples.
The first example is the code of function \code{oneTwo},
a constant, of Fig.~\ref{let-example-graph}:
\begin{curry}
  function oneTwo
    declare x
    declare y
    x = NODE(:, NODE(1), y)
    y = NODE(:, NODE(2), x)
    x[2] = y
    return x
\end{curry}
Symbol application is explicit through \code{NODE}.
The notation \code{x[2]} stands for the second successor
of \code{x}, which is not yet known when the node
referenced by \code{x} is constructed (since the value of \code{y}
is not yet specified at that point).
This allows constructing nodes with a varying number of successors.

The second example is the code of \code{head},
the usual function returning the head of a non-empty list:
\begin{equation}
  \label{head-rules}
  \vbox { \hbox {\hspace*{-2em}
      \begin{curry}
  head (x:_) = x
      \end{curry}
}}
\end{equation}
The rule of \code{head} for the argument \code{[]}
is missing in the Curry source code.
Consequently, the case branch for the argument
\code{[]} is missing in FlatCurry, too.
ICurry has a distinguished expression, \code{exempt},
to capture the absence of a rule:
\begin{curry}
  function head
    declare arg
    arg = ROOT[1]
    case LABEL(arg) of
      [] -> exempt
      :  -> return arg[1]
\end{curry}
where \code{ROOT} is a reference to the root of the expression being
evaluated.  This expression is rooted by \code{head}, which is the reason
why it is passed to function \code{head}.

\subsection{ICurry Use}
\label{ICurry Use}

The stated goal of ICurry is to be a format of Curry programs suitable
for translation into a low-level language. Below, we briefly report
our experience in translating ICurry into various target languages.
Table \ref{size-table} shows the size of a Curry program that translates
ICurry into a target language.
The numerical values in the table, extended from \cite{Wittorf18},
count the lines of code of the translator.
The table is only indicative since ``lines of code''
is not an accurate measure, and
some earlier compilers use older variants of ICurry that have
evolved over time.
\begin{figure}
  \begin{center}
    \renewcommand{\arraystretch}{1.25}
    \setlength{\tabcolsep}{3pt}
    \begin{tabular}{ | l | r | }
      \hline
      C & 441 \\
      \hline
      Python & 342 \\
      \hline
      Java & 790 \\
      \hline
      JavaScript & 632 \\
      \hline 
      JSON & 232 \\
      \hline
    \end{tabular}
  \end{center}
  \caption{\label{size-table}
    Number of Curry source lines of code for various
    translators from ICurry to a target language.
  }
\end{figure}
In all these cases, each ICurry construct has a direct translation
into the target language.
The following details refer to the translation into C.
Declarations and assignment are the same as in C.
The ICurry statements are translated as follows:
(1) the ICurry \emph{return} is the same as in C,
(2) an ICurry \emph{case statement} is translated into a C
\emph{switch statement} where the case selector is the tag of a node,
and (3) the ICurry \emph{exempt} statement is translated into
code that, when executed, terminates the executing computation
without producing any result.
This is justified by the facts that the evaluation strategy executes only
needed steps, and that failures in non-deterministic programs
are natural and expected, therefore they should be silently ignored.

The translation into JSON is used by Sprite \cite{AntoyJost16},
a Curry system under development, whose target language is \cpp.
The JSON format is more convenient when the client of ICurry code
is not coded in Curry.
The ICurry case expressions of a function's
code contain a branch for each constructor in the argument's type
and a branch for each of the following: the choice symbol, the
failure symbol, any function symbol \cite[Fig. 2]{AntoyJost16}, and
any free variable.
A dispatch table, which is addressed by the argument's label's tag, efficiently selects the
branch to be executed.
The behavior of the additional branches is described below, and
is the same across all the functions of a program.
A choice symbol in a pattern matched position results in
the execution of a pull-tabbing step \cite{Antoy11ICLP,BrasselHuchAPLAS07}.
A failure is propagated to the context.
A function symbol triggers the evaluation of the expression
rooted by this symbol.
Finally, a free variable
is instantiated to a choice of \emph{shallow} patterns
of the same type as the variable.
As an example, the evaluation of
\code{head x where x free}, instantiates \code{x}
to \code{[] ? (y:ys) where y, ys free}.
The alternative, \code{[]}, will result in failure.
This can be determined at compile time and removed during optimizations.

A similar approach is described in \cite{Wittorf18} where compilers
from Curry to Python and to C are developed.
As the table above indicates,
the compilers (written in Curry) are quite compact.
We observe that FlatCurry covers the complete language,
since it is the basis for robust Curry implementations,
like PAKCS and KiCS2, and the natural/operational semantics
of Curry are defined in FlatCurry \cite{AlbertHanusHuchOliverVidal05}.
ICurry contains the same information as FlatCurry except type
information, since the type correctness of a program
has been verified at the point of the compilation process
in which ICurry is used.

\subsection{ICurry Generation}
\label{ICurry Generation}

Current Curry distributions such as PAKCS \cite{Hanus18PAKCS} and
KiCS2 \cite{BrasselHanusPeemoellerReck11} provide a package with
the definition of FlatCurry and a rich API for its
construction and manipulation.
Therefore, the ICurry format of a Curry program is conveniently obtained
from the FlatCurry format of that program.

A fundamental difference between the two formats concerns expressions.
Expressions in FlatCurry may contain \emph{cases} and \emph{lets} as
the arguments of a function application.
These are banned in ICurry which allows only nested functional application.
The reason is that the latter can be directly translated into various
imperative languages, where the former cannot.
Therefore, any \emph{case} and \emph{let} constructs that are the arguments
of a function application are replaced by calls to
newly created functions.
These new functions execute these constructs at the top level.
This transformation takes FlatCurry into ICurry,
but it could be executed from FlatCurry into itself,
or even from source Curry into itself.
Our contrived example below shows the latter for ease of understanding.
The code of function \code{g} is irrelevant, therefore, it is not shown:
\begin{curry}
  f x = g x (case x of ...)
\end{curry}
is transformed into:
\begin{curry}
  f x = g x (h x)
  h x = case x of ...
\end{curry}
The offending \emph{case}, as an argument of the application of \code{g}, has been
replaced by a call to a newly created function, \code{h}.
In function \code{h}, the \emph{case} is no longer an argument of a function
application.

The second major difference between FlatCurry and ICurry concerns
case expressions.  FlatCurry matches a selector against
shallow constructor expressions,
where ICurry matches against constructor symbols.
Furthermore, the set of these symbols is complete and ordered in ICurry.
The transformation is relatively simple, except it may require non-local information.
A function in a module $M$ may pattern match on some instance of
a type $t$ that is not declared in $M$.  Therefore, the constructors of $t$
must be accessed in some module different from that being compiled.

A third significant difference between FlatCurry and ICurry concerns
\emph{let blocks}.  They are banned in ICurry, and replaced by
the explicit construction of nodes, and by the assignment of
these nodes' references to local variables.

An algorithm to translate FlatCurry into ICurry is
shown in Appendix~\ref{sec:flatcurry2icurry}.

\section{Related Work and Conclusion}
\label{Related Work and Conclusion}

Our work is centered on the compilation of Curry programs.
As in many compilers, our approach is transformational.
To compile a Curry program $P$, we translate $P$ into a language,
called \emph{target}, for which a compiler already exists.
This is the same route followed
by other Curry compilers like PAKCS \cite{Hanus18PAKCS}
and KiCS2 \cite{BrasselHanusPeemoellerReck11}.

PAKCS translates source Curry code into Prolog, leveraging the
existence of native free variables and non-determinism 
in Prolog.
KiCS2 translates source Curry code into Haskell, leveraging the
existence of first-class function and their efficient
demand-driven execution in Haskell.
Both of these compilers use FlatCurry as an intermediate language.
They have the same front end which translates Curry into FlatCurry.
The use of FlatCurry simplifies the translation process,
but is still appropriate to express Curry computations without much effort .
FlatCurry has some relatively high-level constructs
that can be mapped directly into Prolog and Haskell,
because these languages are high-level too.

Before ICurry, a Curry compiler targeting a C-like language
would handle certain high-level constructs of FlatCurry
in whichever way each programmer would choose.
This led to both duplications of code and unnecessary differences.
ICurry originates from these efforts.
It abstracts the ideas that, over time, proved to be simple and effective
in a language-independent way.
With ICurry, the effort to produce a Curry compiler
targeting an imperative language is both shortened,
because more of the front end can be reused,
and simplified, because the
starting point of the translation is
independent of the target and is well understood.

Our work is complementary to, but independent of, other efforts toward
the compilation of Curry programs.
These efforts include the development
of evaluation strategies \cite{AntoyEchahedHanus00JACM},
or the handling of non-determinism \cite{Antoy11ICLP,BrasselHuchAPLAS07}.

Future work should investigate ICurry to ICurry transformations
that are likely to optimize the generated code.
For example, different orders of the declaration of variables
in a let block lead to different numbers of assignments.
Also, cases expressions as arguments of function
call can be moved outside the call in some situations rather than be
replaced by a call to a new function.

\paragraph{Acknowledgments}
This material is based in part upon work supported 
by the National Science Foundation under Grant No.~1317249.

\bibliographystyle{plain}

\appendix

\section{Translating FlatCurry to ICurry}
\label{sec:flatcurry2icurry}

A \textit{pure expression} is an
expression that only contains literals, variables, constructor
applications, and function applications.  Any $\mathit{or}$ expression
and function application may only contain pure expressions.  The
scrutinee of a case expression must be a variable, literal, or
constructor application.  An assignment in a $\mathit{let}$ expression
must be a pure expression or an $\mathit{or}$ expression.
The branches of a case expression must match all constructors
of a data type in an order fixed by the definition of that data type.
Branches missing in the original Curry program contain $\bot$
in their right-hand side.

The algorithm is divided into five functions which are
described in Fig.~\ref{fig:flatcurryicurry}.
$\mathcal{F}$ translates a FlatCurry function into an ICurry function.
$\mathcal{B}$ translates a FlatCurry expression into an ICurry block.
$\mathcal{D}$ extracts all of the variables declared in a FlatCurry expression.
$\mathcal{A}$ generates necessary assignments for ICurry variables.
$\mathcal{E}$ translates a FlatCurry expression into an ICurry expression.

The functions $\mathcal{F}$ and $\mathcal{E}$ are straightforward
translations.  $\mathcal{F}$ simply makes a block, with the root of the
block being set to the root of the function.  $\mathcal{E}$ is almost
entirely a straight translation, but there is one technical point.  In a
case expression, each branch must be translated into its own block.
However, each of the variables in the pattern of a branch need to be
related to the scrutinee of the case.  This is achieved by setting the
root of the block to the scrutinee of the case.

The $\mathcal{B}$ function creates an ICurry block.
Blocks are more complicated to construct.
Each block will have a root and a list of variables.
The root is the root of the expression that created the block.
For a function, the root is the root of the function expression.
For a case branch, the root is the root of the scrutinee of the case.
The variables of a block are the parameters of a function,
or the pattern variables of a branch.
After declaring variables, all variable in any \emph{let} expressions are declared
with the $\mathcal{D}$ function.
Then each variable is assigned an expression
with the $\mathcal{A}$ function.
If either $\mathcal{D}$ or $\mathcal{A}$ is undefined for some expression,
it does not generate ICurry code.
Finally, we translate the expression into an ICurry statement
with the $\mathcal{E}$ function.

The $\mathcal{D}$ function declares variables declared
in a $\mathit{let}$ or $\mathit{free}$ expression.
If there is a $\mathit{case}$ expression, then a new variable $x_e$ is declared.

The $\mathcal{A}$ function assigns variables in a
$\mathit{let}$ or $\mathit{free}$ expression.
The expression for all variables in a $\mathit{let}$ is
translated with the $\mathcal{E}$ function.  Next, if there are any
variables declared in the $\mathit{let}$ block that are used in one of the
expressions, they need to be filled in.  Finally, if there is a
case expression, we assign $x_e$ to be the root of the scrutinee.

\begin{figure}[t]
\[
\begin{array}{l}
\mathcal{F}(f(x_1,\ldots,x_n) = e) := f = \mathcal{B}(x_1,\ldots, x_n, e, ROOT)\\
\ \\
\mathcal{B}(x_1,\ldots,x_n, \bot, root) := \mathit{exempt} \\
\mathcal{B}(x_1,\ldots,x_n, e, root) := \\
\ \ \mathit{declare }~ x_1\\
\ \ \ldots\\
\ \ \mathit{declare }~ x_n\\
\ \ \mathcal{D}(e)\\
\ \ x_1 = root[1]\\
\ \ \ldots\\
\ \ x_n = root[n]\\
\ \ \mathcal{A}(e)\\
\ \ \mathit{return}~ \mathcal{E}(e)~~~~(\mbox{omit}~ \mathit{return}~ \mbox{if}~ \mathcal{E}(e)~ \mbox{is a}~ \mathit{case})\\
\ \\
\mathcal{D}(\mathit{let}~x_1,\ldots,x_n ~\mathit{free~ in}~ e) := \\
\ \ \mathit{free}~ x_1\\
\ \ \ldots\\
\ \ \mathit{free}~ x_n\\
\mathcal{D}(\mathit{let}~\{x_1 = e_1;\ldots;x_n = e_n\} ~\mathit{in}~ e) :=\\
\ \ \mathit{declare}~ x_1\\
\ \ \ldots\\
\ \ \mathit{declare}~ x_n\\
\mathcal{D}(\mathit{case}~e~\mathit{of}~\{p_1 \to e_1; \ldots; p_n \to e_n\}) := \mathit{declare }~ x_e\\
\ \\
\mathcal{A}(\mathit{let}~\{x_1 = e_1;\ldots;x_n = e_n\} ~\mathit{in}~ e) :=\\
\ \ x_1 = \mathcal{E}(e_1)\\
\ \ \ldots\\
\ \ x_n = \mathcal{E}(e_n)\\
\ \ x_1[p] = x_i ~~~(\mbox{for each occurrence of}~x_i ~\mbox{in}~ e_1 ~\mbox{at position}~ p)\\
\ \ \ldots\\
\ \ x_1[p] = x_i ~~~(\mbox{for each occurrence of}~x_i ~\mbox{in}~ e_n ~\mbox{at position}~ p)\\
\mathcal{A}(\mathit{case}~e~\mathit{of}~\{p_1 \to e_1; \ldots; p_n \to e_n\}) := x_e = \mathcal{E}(e)\\
\ \\
\begin{array}{l@{~:=~}l}
\mathcal{E}(x)                & x\\
\mathcal{E}(c(e_1,\ldots,e_n)) & NODE(c, \mathcal{E}(e_1),\ldots,\mathcal{E}(e_n))\\
\mathcal{E}(f(e_1,\ldots,e_n)) & NODE(f, \mathcal{E}(e_1),\ldots,\mathcal{E}(e_n))\\
\mathcal{E}(e_1~\mathit{or}~e_2) & \mathcal{E}(e_1)~\mathit{or}~\mathcal{E}(e_2)\\
\mathcal{E}(\mathit{let}~\{x_1 = e_1;\ldots;x_n = e_n\} ~\mathit{in}~ e) & \mathcal{E}(e)\\
\mathcal{E}(\mathit{let}~\{x_1,\ldots,x_n\} ~\mathit{free} ~\mathit{in}~ e) & \mathcal{E}(e)\\
\multicolumn{2}{l}{\mathcal{E}(\mathit{case}~e~\mathit{of}~\{c(x_{11},\ldots,x_{1m})\to e_1; \ldots; c(x_{n1},\ldots,x_{nk})\to e_n\}) ~ := } \\
\multicolumn{2}{l}{\ \ 
\begin{array}{ll}
  \mathit{case}~\mathcal{E}(e)~\mathit{of}~\{ & \mathcal{B}(x_{11},\ldots,x_{1m}, e_1, x_e); \\
                                              & \ldots; \\
                                              & \mathcal{B}(x_{n1},\ldots,x_{nk}, e_n, x_e); ~\}
\end{array}}
\end{array}
\end{array}\\
\]
\caption{Algorithm for translating FlatCurry into ICurry}
\label{fig:flatcurryicurry}
\end{figure}

\end{document}